\documentclass[twocolumn,english,floatfix,showpacs]{revtex4}
\usepackage{graphicx}

\makeatletter

\usepackage{amssymb,amsmath}
\usepackage{graphicx}
\newcommand{\mnorm}[1]{%
 \left\vert\kern-0.9pt\left\vert\kern-0.9pt\left\vert #1
 \right\vert\kern-0.9pt\right\vert\kern-0.9pt\right\vert}

\usepackage{babel}
\makeatother
\begin{document}

\title{Optimal quantum chain communication by end gates}

\author{Daniel Burgarth$^{1}$, Vittorio Giovannetti$^{2}$, and Sougato
Bose$^{1}$}

\affiliation{$^{1}$Department of Physics \& Astronomy, University College London,
Gower St., London WC1E 6BT, UK\\
$^{2}$NEST-CNR-INFM \& Scuola Normale Superiore, piazza dei Cavalieri
7, I-56126 Pisa, Italy}

\begin{abstract}
The scalability of solid state quantum computation relies on the ability
of connecting the qubits to the macroscopic world. Quantum chains
can be used as quantum wires to keep regions of external control at
a distance. However even in the absence of external noise their transfer
fidelity is too low to assure reliable connections. We propose a method
of optimizing the fidelity by minimal usage of the available resources,
consisting of applying a suitable sequence of two-qubit gates at the
end of the chain. Our scheme allows also the preparation
of states in the first excitation sector as well as cooling.
\end{abstract}


\maketitle

\section{Introduction}

It is often noted that the advantage of solid state computation is
its \emph{scalability.} This is because a typical chip can contain
a large amount of qubits
and because the fabrication of many qubits is in principle no more
difficult than the fabrication of a single one. In the last couple
of years, remarkable progress was made in experiments with quantum
dots~\cite{QD} and super-conducting qubits~\cite{JJ}. It should
however be emphasized that for initialization, gating and readout,
those qubits have to be connected to the macroscopic world. For
example, in a typical flux qubit gate, microwave pulses are applied
onto specific qubits of the sample. This requires many (classical)
wires on the chip, which is thus a compound of quantum and classical
components. Unfortunately any extra classical control wire is
potentially an independent source of noise as it adds extra coupling
between the quantum computing device and the external world.
Consequently the number of wires is likely to be the bottleneck of
the scalability as a whole: too few will make the device not
powerful enough, too many will make it noisy.

\begin{figure}[htbp]
\begin{center}\includegraphics[%
width=1.0\columnwidth]{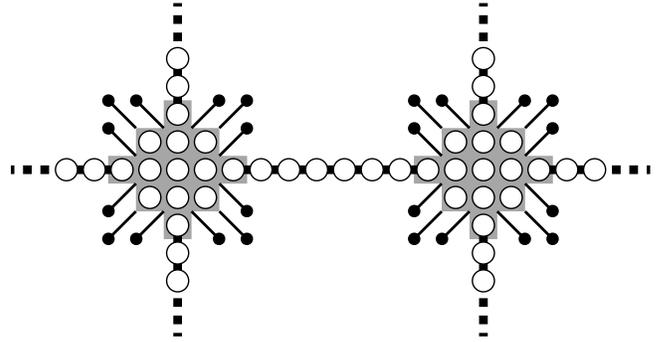}\end{center}
\caption{\label{fig:connect2}
Small blocks (gray) of qubits (white circles)
connected by quantum chains. Each block consists of (say) 13 qubits,
4 of which are connected to outgoing quantum chains (the thick black
lines denote their nearest-neighbor couplings). The blocks are connected
to the macroscopic world through classical wires (thin black lines
with black circles at their ends) through which arbitrary unitary
operations can be triggered on the block qubits. The quantum chains
require no external control. This architecture
is an example of distributed quantum computation~\cite{GROVER} where
the computational  and the communication qubits
are the same objects (i.e. the spins): in this respect no
 interfacing among different qubits species is required
 (compare this with the
implementations of Ref.~\cite{DQC}), whose extreme
difficulty in the context of solid state qubits is discussed for
example, in Ref.\cite{DAMICO}.} 
\end{figure}

In this situation, quantum chains may turn out to be extremely useful
in the development of solid state-based quantum computer technology.
They consist of lines of coupled single qubits \emph{without
external classical control.} In many cases, such permanent couplings
are easy to build in solid state devices. Indeed the really
difficult part usually is to {\em modulate} or to {\em suppress}
them, as has been clearly pointed out for fabricated
hard-wired couplings between super-conducting qubits \cite{zhou} or
tunnel coupled quantum dots \cite{burkard}. Naturally then the
question arises as to whether one can use such quantum chains as
nearly perfect channels for quantum communication despite the lack
of classical controllability. If successful, it will also be the
application of a quantum many-body system for an useful quantum
information processing task.
The setup we have in mind is sketched in Fig.~\ref{fig:connect2}.
It is a distributed quantum computing architecture~\cite{GROVER}
built out of blocks of qubits, some of which are dedicated
to communication and therefore connected to another block through
a quantum chain.
The block size is essentially  determined by the minimum  number of controlling wires necessary
to perform reliable arbitrary  unitary operations on the block spins:
ultimately it depends the ability of
 implementing fault-tolerant gates~\cite{FAULT}
with the available current technology.
The distance between the blocks is instead determined by the length of the
quantum chains between them. It should be large enough to allow for
classical control wiring of each block, but short enough such that
the timescale of the the quantum chain communication is well below
the timescale of decoherence in the system.

Many interesting aspects of quantum chain communication were
investigated in the last years~\cite{ALL,SOU,RAND,HASEL,GAP,ROMITO,RANDSTAB,DAMICO,MEM,LYA1,ERGO,ENG,DECOHERENCE,LYA2},
both from a physics point of view and from a quantum information
point of view. Here, we would like to concentrate on those schemes
\cite{SOU,ENG,GAP,HASEL} which require no further resources than
those outlined in Fig.~\ref{fig:connect2}. The chain couplings may
be engineered \cite{ENG,GAP} to improve the theoretical
communication fidelity, but coupling fluctuations and energy
mismatches will lower the fidelity in
practice~\cite{LYA2,ROMITO,RANDSTAB,DAMICO,RAND}. Hence even without
the contribution of external noise~\cite{DECOHERENCE,RAND} the
quality of transfer may well be too low to yield a scalable system.

In this article we will show that the fidelity can be improved easily
using the gates available in the regions of quantum control. The main
idea is to apply in certain time-intervals two-qubit gates at the
receiving end of the chain. The resulting sequence is determined \emph{a
priori} by the Hamiltonian of the system. As we shall see, the maximal
fidelity that can be reached this way is limited only by external
noise, and not by the spatial fluctuations of the couplings (cf. \cite{RAND}).
This is similar in spirit to the dual-rail \cite{RAND} and memory
protocols \cite{MEM}, but here we give a protocol that is \emph{optimal}
in the resources used: a single spin chain and a two-qubit gate at
the each end. It is optimal because two-qubit gates at the sending
and receiving end are required in order to connect the chain to the
blocks in \emph{all} above protocols (though often not mentioned explicitly).
Our scheme has some similarities with \cite{HASEL}, but the gates
used here are much simpler, and arbitrarily high fidelity is guaranteed
by a convergence theorem for arbitrary coupling strengths and all
non-Ising coupling types that conserve the number of excitations.
Furthermore, we show numerically that our protocol could also be realized
by a simple switchable interaction. This means that quantum state
transfer experiments with our protocol could be performed well before
the realization of the blocks from Fig.~\ref{fig:connect2}.

The paper is organized as follows. In Sec.~\ref{arbitrary}
we introduce the protocol and we present an analytical proof of the 
asymptotical convergence of the associated transfer fidelity. 
To do so we restrict ourself to a regime in which the two-qubit
gates applied at the end of the chain act instantaneously, i.e. they 
are activated over 
time intervals  which are much shorter than  
the typical time scale of the free spin evolution.
This hypothesis is not fundamental but it allows us to simplify
the math: we will drop it in Sec.~\ref{practical} where, 
by using numerical techniques, we  generalize the convergence analysis 
to cases in which the timing of the end gates 
are
comparable with those of the free dynamical evolution of the chain.  
The manuscript ends with the conclusions in Sec.~\ref{conclusion}.

\section{Arbitrarily Perfect State Transfer}\label{arbitrary}
\begin{figure}[t]
\begin{center}\includegraphics[%
  width=1.0\columnwidth]{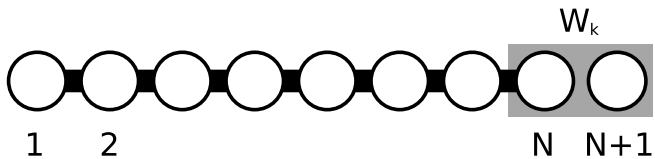}\end{center}
\caption{\label{fig:setup}A quantum chain (qubits $1,2,\cdots,N$) and a
target qubit ($N+1$). By applying a sequence of two-qubit unitary
gates $W_{k}$ on the last qubit of the chain and the target qubit,
arbitrarily high fidelity can be achieved for the transmission of quantum
information from the left hand side to the right hand side of the chain.}
\end{figure}

Here we present an analytical proof of the 
convergence of our transferring protocol. 
For the sake of simplicity we will focus 
on a single chain from the setup of Fig.~\ref{fig:connect2}.
In this case, as in Ref.~\cite{SOU},
 the left end side of the chain 
plays effectively the role of
a sender of quantum information while the right end side 
plays the role of a receiver.
Within this framework we will show that 
the receiving block (gray area of Fig.~\ref{fig:setup}) 
can improve the transmission fidelity to an arbitrarily
high value by applying suitable two-qubit gates $W_k$ (see below)
between the end of the chain
and a {}``target qubit'' of the block.  As mentioned in the introduction,
in order to get analytical results, we will restrict the analysis to the
case in which the gates $W_k$ act instantaneously on the system.

\subsection{Notation}
Before entering into the details of the derivation let fix some notation
and define the property of the system.
We label the qubits of the
chain by $1,2,\cdots,N$ and the target qubit by $N+1.$ We also define
the states\begin{eqnarray*}
|\mathbf{0}\rangle & \equiv & |00\ldots0\rangle\\
|\mathbf{n}\rangle & \equiv & \sigma_{n}^{+}|\mathbf{0}\rangle\quad n=1,2,\ldots,N+1,\end{eqnarray*}
where $\sigma_{n}^{+}$ is the Pauli $\sigma^{+}$ operator acting
on the $n$th qubit. 
With these definitions the typical initial configuration of our
communication protocol will be described by vectors of the
form 
\begin{eqnarray}
|\psi_{\textrm{initial}}\rangle=\alpha|\mathbf{0}\rangle+\beta|\mathbf{1}\rangle \;,
\label{initial}
\end{eqnarray}
where all the qubits from $2$ to $N+1$ 
are in the reference state $|0\rangle$ while
the first qubit has being prepared into the logical state $\alpha
|0\rangle + \beta |1\rangle$. This the quantum bits that one would like
to propagate along the chain.

The free evolution of the system is
described by a Hamiltonian $H$ which couples all the qubits but
the target.
Our main assumption  on  
$H$ is that it has $|\mathbf{0}\rangle$ as eigenstate with
eigenvalue $0$, i.e. $H|\mathbf{0}\rangle=0$,
 and 
a $N$ dimensional invariant subspace
spanned by the vectors  $\left\{ |\mathbf{n}\rangle;\; n=1,2,\ldots,N\right\}$.
Under this condition $H$ corresponds to a Hamiltonian that conserves the
number of excitations along the chain, which would be the case for
example of the Heisenberg or XY chains  considered in most of the
protocol proposed so far~\cite{ALL,SOU,ENG,HASEL,RAND,MEM}. 
Thanks to this property the analysis of the protocol can be restricted
to the $N+2$ dimensional Hilbert 
$\mathcal{H}=\textrm{span}\left\{ |\mathbf{n}\rangle;\; n=0,1,2,\ldots,N+1\right\}$.
Our final assumption about $H$ 
is that there exists a time $t$ such that
$\langle\mathbf{N}|\exp\left\{ -itH\right\} |\mathbf{1}\rangle\neq0$.
Physically this means that the Hamiltonian has the capability of transporting excitations (and hence information)
from the first to the last qubit of the chain. As mentioned in the
introduction, the fidelity of this transport may be very bad in practice.

We denote the unitary evolution operator for a given time $t_{k}$
as $U_{k}\equiv\exp\left\{ -it_{k}H\right\} $ and introduce the projector\[
P=\openone_{\mathcal{H}}-|\mathbf{0}\rangle\langle\mathbf{0}|-|\mathbf{N}\rangle\langle\mathbf{N}|-|\mathbf{N+1}\rangle\langle\mathbf{N+1}|.\]
A crucial ingredient to our protocol is the unitary transformation
\begin{eqnarray}
W(c,d) & \equiv P+ & |\mathbf{0}\rangle\langle\mathbf{0}|+d|\mathbf{N}\rangle\langle\mathbf{N}|+d^{*}|\mathbf{N+1}\rangle\langle\mathbf{N+1}|
\nonumber \\
 &  & +c^{*}|\mathbf{N+1}\rangle\langle\mathbf{N}|-c|\mathbf{N}\rangle\langle\mathbf{N+1}|,\label{crucial} 
\end{eqnarray}
where $c$ and $d$ are complex normalized amplitudes. 
One can easily verify that $W$  acts as the
identity on all but the last two qubits, and can hence be realized
by \emph{a local two-qubit gate on the qubits $N$ and $N+1$.} Furthermore
we have $WP=P$ and\begin{equation}
W(c,d)\left[\left\{ c|\mathbf{N}\rangle+d|\mathbf{N+1}\rangle\right\} \right]=|\mathbf{N+1}\rangle.\label{eq:wdesign}\end{equation}
The operator $W(c,d)$ has the role of moving probability amplitude
$c$ from the $N$th qubit to target qubit. It can be applied locally
by the receiving block.

\subsection{The protocol}
Using the time-evolution operator $U_k$ and two-qubit unitary gates  on the
qubits $N$ and $N+1$ defined in Eq.~(\ref{crucial})
we will now develop a protocol that transforms
the state $|\mathbf{1}\rangle$ into $|\mathbf{N+1}\rangle.$ Let
us first look at the action of $U_{1}$ on $|\mathbf{1}\rangle.$
Using the projector $P$ we can decompose this time-evolved state
as\begin{eqnarray} 
U_{1}|\mathbf{1}\rangle & = & PU_{1}|\mathbf{1}\rangle+|\mathbf{N}\rangle\langle\mathbf{N}|U_{1}|\mathbf{1}\rangle \label{evolved}\\
 & \equiv & PU_{1}|\mathbf{1}\rangle+\sqrt{p_{1}}\left\{ c_{1}|\mathbf{N}\rangle+d_{1}|\mathbf{N+1}\rangle\right\} , \nonumber \end{eqnarray}
where $d_1=0$ and 
\begin{eqnarray}
p_{1}=\left|\langle\mathbf{N}|U_{1}|\mathbf{1}\rangle\right|^{2}\;, \quad 
c_{1}=\langle\mathbf{N}|U_{1}|\mathbf{1}\rangle/\sqrt{p_{1}}\;. \nonumber 
\end{eqnarray}
 Let us now consider the instantaneous application  of 
 the unitary transformation $W_{1}\equiv W(c_{1},d_{1})$
on the time-evolved state of Eq.~(\ref{evolved}). According to
 Eq.~(\ref{eq:wdesign}) this yields 
\begin{eqnarray}
W_{1}U_{1}|\mathbf{1}\rangle & = & PU_{1}|\mathbf{1}\rangle+\sqrt{p_{1}}|\mathbf{N}+1\rangle.\label{eq:firsttime}\end{eqnarray}
Hence with a probability of $p_{1},$ the excitation is now in the
position $N+1,$ where it is {}``frozen'' (that qubit is not
coupled to the chain). We will now show that at the next step, this
probability is increased. Applying $U_{2}$ to Eq. (\ref{eq:firsttime})
we get \begin{eqnarray*}
\lefteqn{U_{2}W_{1}U_{1}|\mathbf{1}\rangle}\\
 & = & PU_{2}PU_{1}|\mathbf{1}\rangle+\langle\mathbf{N}|U_{2}PU_{1}|\mathbf{1}\rangle|\mathbf{N}\rangle+\sqrt{p_{1}}|\mathbf{N}+1\rangle\\
 & = & PU_{2}PU_{1}|\mathbf{1}\rangle+\sqrt{p_{2}}\left\{ c_{2}|\mathbf{N}\rangle+d_{2}|\mathbf{N}+1\rangle\right\} \end{eqnarray*}
with 
\begin{eqnarray}
c_{2}&=&\langle\mathbf{N}|U_{2}PU_{1}|\mathbf{1}\rangle/\sqrt{p_{2}}\;, 
\qquad d_{2}=\sqrt{p_{1}}/\sqrt{p_{2}} \;, \\
p_{2} & = & p_{1}+\left|\langle\mathbf{N}|U_{2}PU_{1}|\mathbf{1}\rangle\right|^{2}\ge p_{1}.\end{eqnarray}
Applying $W_{2}\equiv W(c_{2},d_{2})$ we get \[
W_{2}U_{2}W_{1}U_{1}|\mathbf{1}\rangle=PU_{2}PU_{1}|\mathbf{1}\rangle+\sqrt{p_{2}}|\mathbf{N}+1\rangle.\]
Repeating this strategy $\ell$ times we get\begin{equation}
\left(\prod_{k=1}^{\ell}W_{k}U_{k}\right)|\mathbf{1}\rangle=\left(\prod_{k=1}^{\ell}PU_{k}\right)|\mathbf{1}\rangle+\sqrt{p_{\ell}}|\mathbf{N}+1\rangle,\label{eq:general}\end{equation}
where the products are arranged in the time-ordered way. Using the
normalization of the r.h.s. of Eq. (\ref{eq:general}) we get\[
p_{\ell}=1-\left\Vert \left(\prod_{k=1}^{\ell}PU_{k}\right)|\mathbf{1}\rangle\right\Vert ^{2}.\]
From Ref. \cite{RAND} we know that there exists a $\tau>0$ such
that for equal time intervals $t_{1}=t_{2}=\ldots=t_{k}=\tau$ we
have $\lim_{\ell\rightarrow\infty}p_{\ell}=1.$ Therefore the limit
of infinite gate operations for Eq. (\ref{eq:general}) is given by\begin{equation}
\lim_{\ell\rightarrow\infty}\left(\prod_{k=1}^{\ell}W_{k}U_{k}\right)|\mathbf{1}\rangle=|\mathbf{N+1}\rangle.\label{eq:convergence}\end{equation}
It is also easy to see that 
\begin{eqnarray}
\lim_{k\rightarrow\infty}d_{\ell}=1 \;,  \qquad
\lim_{k\rightarrow\infty}c_{\ell}=0 \;,
\end{eqnarray}
which shows that for large $k$ the gates $W_k$ converge
to the identity operator, i.e. 
$\lim_{k\rightarrow\infty}W_{k}=\openone_{\mathcal{H}}.$

Equation~(\ref{eq:convergence}) is the main result of the paper.
Together with the fact that $W_{k}U_{k}$ leaves the vector 
$|\mathbf{0}\rangle$
invariant (i.e. $W_{k}U_{k}|\mathbf{0}\rangle=|\mathbf{0}\rangle$),
this expression can be used to show that 
an arbitrary and unknown qubit at the first site~(\ref{initial}) 
is transferred to the last site, i.e. 
\begin{eqnarray}
|\psi_{\textrm{initial}}\rangle \longrightarrow 
|\psi_{\textrm{final}}\rangle=\alpha|\mathbf{0}\rangle+\beta|\mathbf{N+1}\rangle\;, \end{eqnarray}
This corresponds to an arbitrarily perfect state transfer. As discussed
in \cite{ERGO}, the convergence of Eq.~(\ref{eq:convergence}) 
is asymptotically exponentially
fast in the number of gate applied (a detailed analysis of the relevant
scaling can be found in~\cite{RAND}). 
Equation (\ref{eq:convergence})
shows that \emph{any non-perfect transfer
can be made arbitrarily perfect} by only applying two-qubit gates
on one end of the quantum chain. On one hand, it avoids restricting the gate times
to specific times (as opposed to the scheme in \cite{RAND}) while
requiring no additional memory qubit (as opposed to the scheme in
\cite{MEM}). On the other hand, given the similarities with the convergence proof 
of the protocols of Ref. [17], the speed of the convergence of the present scheme
is expected to be similar to that associated with such protocols.
It is worth noticing that the sequence of unitary transformations 
$W_{k} \equiv W(c_k,d_k)$ that needs to be applied to the end of the chain
to perform the state transfer is only depending on the Hamiltonian $H$ 
of the quantum chain. The relevant properties can in principle be
determined a priori by preceding measurements and tomography on the
quantum chain (as discussed in Ref.~\cite{RAND}). 
Furthermore, by performing
projective measurements and conditional spin flips on the memory qubit instead,
the chain can also be \emph{cooled} (this follows from the convergence theorem in higher
excitation sectors given in~\cite{MEM,YUASA}). Even state preparation of arbitrary
known states in the first excitation sector is possible by using a time-inversed protocol~\cite{BB}.

\begin{figure}[t]
\begin{center}\includegraphics[%
  width=1.0\columnwidth]{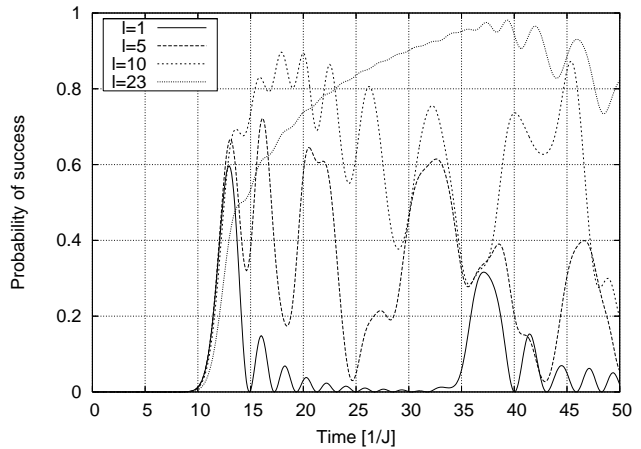}\end{center}
\caption{\label{fig:optimization}Even with a finite number $\ell$ of two-qubit operations,
the success probability of the transfer can be improved significally. We give a numerical
example of a Heisenberg chain of length $N=23$, where the gate times are equidistant. 
In particular, the plots show the transfer fidelity achievable after 
a time $t$ has been elapsed from
the initial condition assuming that in the time interval $[0,t]$, 
$\ell$ two-qubit operations $W_k$ have being performed
at times $t_1 = t/\ell$, $t_2 = 2t/\ell$, $\cdots$, $t_{\ell-1} = (\ell -1)t/\ell$ and $t_\ell=t$. 
For $\ell=1$ only a single two-qubit gate is performed to transfer the information and
 our result coincides
with the original protocol~\cite{SOU}. Already for $\ell=10$ we find an improvement of approx. 50\% within the same time-scale. For $\ell=N$ we obtain a near-perfect transfer. Notice that
starting to extract information from the chain too early causes a small quantum Zeno effect
(e.g. see  the case $\ell =23$ which for $t\sim 10/J$  is outperformed by
the original protocol~\cite{SOU}).}
\end{figure}

Of course,  even though 
perfect quantum state transfer is achieved only
in the limit of infinitely many steps,
nothing prohibits one to stop the protocol 
after a finite number of applications 
of $W_k$. In this case the resulting communication fidelity will not be 
optimal but will be still higher than that obtained
in those scheme which only exploit direct propagation of the 
excitations along the chain~\cite{SOU} (see Fig.~\ref{fig:optimization}).
In realistic scenarios the choice of the  maximum 
number of steps one can 
use will depend upon the presence of external
noise sources that determine
 the coherence time-scales of the system. Since our scheme applies to all Hamiltonians
that conserve the number of excitations, it can also be applied to improve the schemes
that use engineered couplings~\cite{GAP,ENG} in the presence of disorder. In this situation the initial fidelity is already quite high, and the required number of operations is even lower.

\section{Generalization}\label{practical}

Motivated by the  result of
the previous sections
 we now investigate how the protocol
may be implemented in practice, well before the realization \emph{}of
the quantum computing blocks from Fig.~\ref{fig:connect2}. 

The two-qubit
gates $W_{k}$ are essentially rotations in the $\{|01\rangle,|10\rangle\}$
space of the qubits $N$ and $N+1.$ It is therefore to be expected
that they can be realized (up to a irrelevant phase) by a switchable
Heisenberg or $XY$ type coupling between the $Nth$ and the target
qubit. However in the above, we have assumed that the gates $W_{k}$
can be applied instantaneously, i.e. in a time-scale much smaller
than the time-scale of the dynamics of the chain. This corresponds
to a switchable coupling that is much stronger than the coupling strength
of the chain. %
\begin{figure}[t]
\begin{center}\includegraphics[%
  width=1.0\columnwidth]{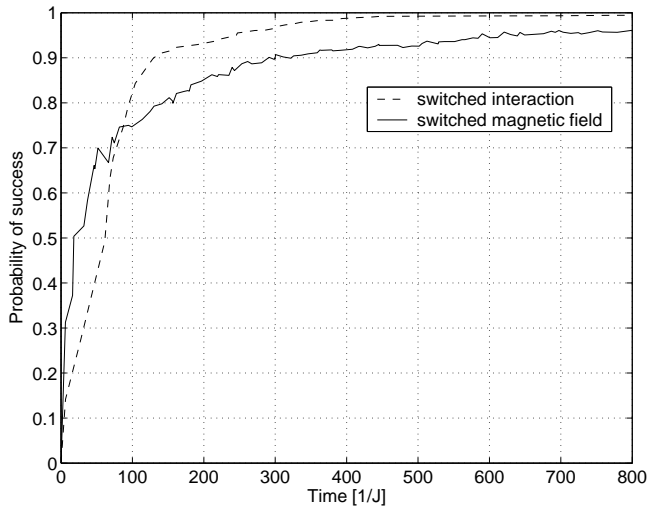}\end{center}
\caption{\label{fig:numerics}Numerical example for the convergence of the
success probability. Simulated is a quantum chain of length $N=20$
with the Hamiltonian from Eq. (\ref{eq:heis}) (dashed line) and Eq.
(\ref{eq:mag}) with $B/J=20$ (solid line). Using the original protocol
\cite{SOU}, the same chain would only reach a success probability
of $0.63$ in the above time interval.}
\end{figure}
 Here, we numerically investigate if a convergence similar to the
above results is still possible when this assumption is not valid.
We \emph{do} however assume that the switching of the interaction
is still describable by an instantaneous switching (i.e. the sudden
approximation is valid). This assumption is mainly made to keep the
numerics simple. We do not expect qualitative differences when the
switching times become finite as long as the time-dependent Hamiltonian
is still conserving the number of excitations in the chain. In fact
it has recently been shown that the finite switching time can even
\emph{improve} the fidelity \cite{LYA2}.

We have investigated two types of switching. For the first type, the
coupling itself is switchable, i.e.\begin{equation}
H(t)=J\sum_{n=1}^{N-1}\sigma_{n}^{-}\sigma_{n+1}^{+}+\Delta(t)\sigma_{N}^{-}\sigma_{N+1}^{+}+\textrm{h.c.},\label{eq:heis}\end{equation}
where $\Delta(t)$ can be $0$ or $1.$ For the second type, the target
qubit is \emph{permanently} coupled to the remainder of the chain,
but a strong magnetic field on the last qubit can be switched, \begin{equation}
H(t)=J\sum_{n=1}^{N}\sigma_{n}^{-}\sigma_{n+1}^{+}+\textrm{h.c.}+B\Delta(t)\sigma_{N+1}^{z},\label{eq:mag}\end{equation}
where again $\Delta(t)$ can be $0$ or $1$ and $B\gg1.$ This suppresses
the coupling between the $N$th and $N+1$th qubit due to an energy
mismatch.

For the purposes
of the present discussion it is sufficient to
focus on a specific choice of control pulses $\Delta(t)$: this will not give
us the best achievable performances but it will prove our point.
Therefore in both cases, we first numerically optimize the times for unitary
evolution $t_{k}$ over a fixed time interval such that the probability
amplitude at the $N$th qubit is maximal.
The algorithm then finds
the optimal time interval during which $\Delta(t)=1$ such that the
probability amplitude at the target qubit is increased. In some cases
the phases are not correct, and switching on the interaction would
result in probability amplitude floating back into the chain. In this
situation, the target qubit is left decoupled and the chain is evolved
to the next amplitude maximum at the $N$th qubit. Surprisingly, even
when the time-scale of the gates is comparable to the dynamics, near-perfect
transfer remains possible~(Fig \ref{fig:numerics}). In the case
of the switched magnetic field, the achievable fidelity depends on
the strength of the applied field. This is because the magnetic field
does not fully suppress the coupling between the two last qubits.
A small amount of probability amplitude is lost during each time evolution
$U_{k},$ and when the gain by the gate is compensated by this loss,
the total success probability no longer increases.

\section{Conclusions}\label{conclusion}

We found an optimal strategy for achieving arbitrarily perfect state
transfer and state preparation (including cooling) by applying a sequence
of two-qubit operations at the receiving end of a quantum chain.
Surprisingly, the gates can be realized by a switchable interaction
of the same strength as the chain coupling. By pointing out
the rather counterintuitive fact that minimal control at one end
enables a large class of quantum many-body systems to be used as a
perfect quantum wire, we open up the field of as to whether a
similar result holds for other many-body systems. 
\acknowledgments
We would like to
thank Floor Pauw, Andriy Lyakhov, Christoph Bruder and Rosario
Fazio for stimulating discussions. DB acknowledges the support of
the UK Engineering and Physical Sciences Research Council, Grant Nr.
GR/S62796/01. VG is grateful to SB for the hospitality at UCL.


\begin{thebibliography}{10}
\bibitem{QD}F. H. L. Koppens et. al., Science \textbf{309}, 1346-1350 (2005);
R. Hanson et. al., Phys. Rev. Lett. \textbf{94}, 196802 (2005).
\bibitem{JJ}T. Yamamoto et. al., Nature \textbf{425}, 941 (2003); I. Chiorescu
et. al., Nature \textbf{431},159 (2004).
\bibitem{zhou} X. Zhou {\em et. al.}, Phys. Rev. Lett. {\bf 89}, 197903
(2002).
\bibitem{burkard} R. Hanson and G. Burkard, Phys. Rev. Lett. {\bf 98}, 050502
(2007).
\bibitem{GROVER} L. Grover, quant-ph/9704012;  R. Von Meter, K. Nemoto, and W. J. Munro, quant-ph/0701043.
\bibitem{FAULT} P. Shor,  Proc. 35th Annu. Symp. Fundamentals of Computer Science
{\bf 124} (IEEE Press, Los Alamitos, CA, 1994).
\bibitem{DQC}J. I. Cirac, {\em et al.} Phys. Rev. A {\bf 59} (1999);
A. M. Steane and D. M. Lucas, Fortschritte der Physik, {\bf 48} 839 (2000);
D. K. L. Oi, S. J. Devitt, and L. C. L. Hollenberg, Phys. Rev. A {\bf 74} 052313 (2006);
A. Serafini, S. Mancini, and S. Bose, Phys. Rev. Lett. {\bf 96} 010503 (2006).
\bibitem{ALL}J. Eisert et. al., Phys. Rev. Lett. \textbf{93} 190402 (2004); M.
B. Plenio et. al., New J. Phys. \textbf{6}, 36 (2004); L. Amico et.
al., Phys. Rev. A. \textbf{69}, 022304 (2004); T. J. Osborne and N.
Linden, Phys. Rev. A \textbf{69}, 052315 (2004); Y. Li et. al., Phys.
Rev. A \textbf{71}, 022301 (2005); V. Giovannetti and R. Fazio, Phys.
Rev. A \textbf{71}, 032314 (2005); M. Paternostro et al., Phys. Rev.
A \textbf{71}, 042311 (2005); A. Bayat and V. Karimipour, Phys. Rev.
A \textbf{71}, 042330 (2005); T. Boness et. al., Phys. Rev. Lett.
\textbf{96}, 187201 (2006); M. J. Hartmann et al., New J. Phys. \textbf{8}
(2006) 94; O. M{\"u}lken and A. Blumen, Phys. Rev. A \textbf{73}, 012105
(2006); J. Zhang et. al., Phys. Rev. A \textbf{73}, 062325 (2006);
S. Paganelli et. al., Phys. Rev. A \textbf{74}, 012316 (2006); M.
Avellino et. al., Phys. Rev. A \textbf{74}, 012321 (2006); M. H. Yung,
Phys. Rev. A \textbf{74} 030303(R) (2006); J. Fitzsimons et. al., quant-ph/0606188;
D. Rossini et. al., quant-ph/0609022.
\bibitem{SOU}S. Bose, Phys. Rev. Lett. \textbf{91}, 207901 (2003).
\bibitem{GAP}M. B. Plenio and F. L. Semiao, New. J. Phys. \textbf{7}, 73 (2005);
A. Wojcik et al., Phys. Rev. A \textbf{72}, 034303 (2005); A. Wojcik
et. al., quant-ph/0608107
\bibitem{ENG}M. Christandl et. al., Phys. Rev. Lett. \textbf{92}, 187902 (2004);
C. Albanese et. al., Phys. Rev. Lett. \textbf{93}, 230502 (2004);
G. M. Nikolopoulos et. al., Europhys. Lett. \textbf{65}, 297 (2004);
G. M. Nikolopoulos et al., J. Phys.:Condens. Matter \textbf{16}, 4991
(2004); M. Christandl et. al., Phys. Rev. A \textbf{71}, 032312 (2005);
P. Karbach and J. Stolze, Phys. Rev. A \textbf{72}, 030301(R) (2005);
A. Kay, Phys. Rev. A \textbf{73}, 032306 (2006).
\bibitem{HASEL}H. L. Haselgrove, Phys. Rev. A \textbf{72}, 062326 (2005).
\bibitem{ROMITO}A. Romito et al., Phys. Rev. B \textbf{71}, 100501(R) (2005).
\bibitem{RANDSTAB}G. De Chiara et. al., Phys. Rev. A \textbf{72}, 012323 (2005).
\bibitem{DAMICO}Irene D'Amico, Microelectronics Journal {\bf 37}, 1440 (2006).
\bibitem{LYA2}A. O. Lyakhov and C. Bruder, Phys. Rev. B \textbf{74}, 235303 (2006).
\bibitem{RAND}D. Burgarth and S. Bose, Phys. Rev. A \textbf{71}, 052315 (2005);
D. Burgarth, V. Giovannetti, and S. Bose, J. Phys. A: Math. Gen. \textbf{38}
6793 (2005).D. Burgarth and S. Bose, New J. Phys. \textbf{7} 135 (2005).
\bibitem{DECOHERENCE}D. Burgarth and S. Bose, Phys. Rev. A \textbf{73}, 062321 (2006);
JM Cai et. al., Phys. Rev. A \textbf{74}, \textbf{}022328 (2006);
L. Zhou et. al., quant-ph/0608135.
\bibitem{MEM}V. Giovannetti and D. Burgarth, Phys. Rev. Lett. \textbf{96}, 030501
(2006).
\bibitem{LYA1}A. Lyakhov and C. Bruder, New J. Phys. \textbf{7}, 181 (2005).
\bibitem{ERGO}D. Burgarth and V. Giovannetti, New J. Phys. \textbf{9}, 150 (2007).

\bibitem{YUASA} H.Nakazato, H. et. al., Phys. Rev. A \textbf{70}, 012303 (2004).
\bibitem{BB} D. Burgarth and V. Giovannetti, arXiv:0704.3027.

\end{thebibliography}
\end{document}